# Novel approach for non-invasive phase-sensitive 2D imaging of biological objects with photovoltaic trapping

Lusine Tsarukyan[*], Anahit Badalyan, and Rafael Drampyan

*Institute for Physical Research, National Academy of Sciences of Armenia, 0204, Ashtarak-2, Armenia*

## ABSTRACT

A novel *non-invasive* microscopy technique for imaging and sizing of folded DNA molecules with the use of photovoltaic tweezers and phase-sensitive detection is elaborated and realized. This novel method is compared with the state-of-the-art method of visualization of DNA molecules by fluorescent microscopy technique which requires the labeling of the DNA by dye molecules. The advantage of the novel method is no requirement for the incubation of DNA with a fluorescent label, as well as the simultaneous observation of the light-induced inside the crystal refractive lattice (i.e. photovoltaic field configuration) and the trapped micro-objects by the phase-contrast method. The suggested novel method provides the accurate measurement of the size and shape of the folded DNA molecules as well as the determination of the charged or neutral nature of the trapped bio-objects by their disposition relative to the lattice fringes. We demonstrate that the labeling of DNA causes an overestimation of the molecular size. The method can be extended to the other bio-objects. The high-performance photovoltaic tweezers operating in an autonomous regime as *lab-on-a-chip* devices are promising for applications in integrated optics, microscopy, micro- and nanoelectronics, and biotechnology.

**Keywords:** Microparticle imaging; photovoltaic tweezers; lithium niobate; Bessel beam; DNA; phase sensitive microscopy.

## 1. Introduction

Elaboration of the tweezers for trapping and manipulation of micro- and nanoparticles is a challenging task in many research fields. They successfully can be used in photonic devices, micro- and nanoelectronics, optical microscopy, and biotechnology. Very promising are the laser-assisted methods for manipulation and trapping of micro- and nanoparticles.

Optical tweezers are well-established tools [1] for trapping micro- and nanoparticles by focused light beams. Neutral particles also can be manipulated electro-kinetically via dielectrophoretic (DEP) forces exerted by inhomogeneous static electric fields provided by electrodes [2, 3]. Dielectrophoresis has been proven as a powerful method for micromanipulation and separation of biological objects [4-9]. However, the method lacks flexibility because of the static electrodes.

The combination of optical and electro-kinetic methods is possible by the use of photorefractive crystals. The essence of the effect relates to optical holography [10, 11]. The spatially inhomogeneous illumination of photorefractive crystals redistributes photo-charges, excited by photoionization and generates the alternating space-charge electric fields ($\sim 10^7$ V/m) in the volume and *proximity* of the crystal surface associated with the bulk photovoltaic (PV) effect [12, 13]. The PV fields induced in the photorefractive crystal modulate the refractive index via the electro-optic effect ($\delta n \sim E$) thus creating the refractive structures inside the crystal [11].

---

[*] Corresponding author.
*E-mail address*: lustsarukyan@gmsail.com (L. Tsarukyan)

The *E*-field near the crystal surface can act on micro- or nano-objects disposed close to the crystal surface and provide the trapping and manipulation of neutral micro- and nanoparticles on the surface of the crystal via DEP forces proportional to the *E*-field intensity gradient ($F_{dep} \sim \mathrm{grad} E^2$). Such trapping devices are called *PV tweezers* [14].

The PV evanescent electric fields *E* near the crystal surface induce electrical forces on either the particles that have electrical charge *q* (electrophoretic forces $F_{EP} = qE$) or the neutral particles that become polarized by an electric field (DEP forces).

The DEP force $F_{DEP}$ which affects the neutral micro- and nanoparticles of spherical form with a dielectric constant $\varepsilon_p$ in the transparent surrounding medium with a dielectric constant $\varepsilon_m$ under nonuniform static electric field *E* is given by expression [2, 3, 12, 13]:

$$F_{DEP} = 2\pi\varepsilon_m R^3 K \nabla E^2, \quad K = \frac{\varepsilon_p - \varepsilon_m}{\varepsilon_p + 2\varepsilon_m} \qquad (1)$$

where *R* is the particle radius. The sign and magnitude of the DEP force are dependent on the Clausius-Mossotti factor *K* [2, 12].

The DEP force depends on the particle volume and gradient of the scalar value $E^2$. When the particle dielectric constant $\varepsilon_p$ is larger than THE dielectric constant $\varepsilon_m$ of a medium, the force is attractive, while the force becomes repulsive in the opposite case.

In most experiments, the inhomogeneous illumination of the photorefractive crystal was performed by the interference of two laser beams providing the formation of nearly sinusoidal one-dimensional (1D) light patterns with periodicities of a few tens of micrometers [12,13]. The trapped particles (dielectric, metallic) were detected by a reflective microscope. The recent achievements in the use of PV tweezers for biological research (the tumor cell, the alive *E-coli* bacteria, and the spores and pollen grains are presented in [14-18].

The application of the nondiffracting Bessel beam technique for the elaboration of 2D, high-contrast PV tweezers was suggested and realized in our recent work [19] for the trapping of dielectric and metal microparticles. Bessel beam provides an extremely high *E*-field gradient, due to its large modulation depth of m = 1. The advantages of the elaborated Bessel beam-assisted PV tweezers are the high-performance operation, and very long lifetime (up to one year) of the PV field due to the high resistance of the used Fe-doped lithium niobate crystal (LN:Fe) for suitable Fe ions concentrations [20], thus providing the operation of chip-scale tweezers in *autonomous regime* (*lab-on-a-chip* device).

In [19] the trapped Ag metallic nanoparticles were observed by a reflective microscope because of the high reflectance of Ag particles. The first results on the trapping of DNA molecules in NaCl salt solutions by PV tweezers and their observation with the use of phase-sensitive polarization microscopy technique were reported in [21].

The state-of-the-art method for imaging the folded DNA molecules is the fluorescence microscopy technique (see [22] and Refs. therein). The DNA molecules, which previously have been incubated with selective DNA fluorescent labels, like ethidium bromide or DAPI, are observed by fluorescent microscope. The labeled by dye DNA molecules are visualized by fluorescence under UV excitation. The drawback of the method is the alteration of the structure, shape, and size of the labeled DNA molecule.

The elaboration of label-free and non-invasive techniques for the study of bio-objects is a challenging task. The well-known non-invasive technique is, for example, second harmonic generation (SHG) microscopy widely used for the study of tissue [23, 24]. Recently [25] it was demonstrated that SHG-technique can be applied to dehydrated A-type DNA. The method is expansive as requires the frequency tunable TiSa femtosecond laser and sensitive registration as the conversion efficiency to the second harmonic radiation is small $\sim 10^{-5}$.

We are suggesting a new label-free and non-invasive approach for the 2D imaging of DNA molecules by combining the PV trapping of the DNA molecules on the LN:Fe crystal surface and simultaneous observation of both refractive lattice (i.e. PV field distribution) light-induced in the crystal and the trapped molecules by phase-sensitive polarization microscopy technique. The advantage of the novel method is no requirements for incubation of DNA with fluorescent labels. The method is simple and provides an accurate measurement of the size and shape of any type of DNA molecule.

The comparative studies of the elaborated non-invasive phase-sensitive PV trapping technique and the state-of-the-art fluorescence microscopy technique for imaging the folded DNA molecules which remained open are the main aim and novelty of our studies. Our novel finding is that the labeling of DNA causes an overestimation of the molecular size.

The method can be extended to other bio-objects. The suggested novel method based on PV tweezers also allows the determination of the charged or neutral nature of the trapped bio-objects by their disposition relative to the lattice fringes.

## 2. Materials and methods

The following scientific methods were used in the experiment: (i) the method of formation of Bessel beam and recording of Bessel lattice in LN:Fe crystal, (ii) preparing of photorefractive LN crystals doped with Fe impurity ions providing excellent photorefractive properties (iii) DNA sample preparation, and (iv) the detection of the trapped micro- and nanoparticles.

(i) *Bessel beam.* Among different methods of holographic recording of the photonic lattices, the *non-diffracting* beams [26] are of special interest. Non-diffracting beams have the feature of conserving their transverse intensity distribution during propagation in free space. This feature makes diffraction-free beams very promising for the formation of high-contrast photonic structures and space charge fields in the photorefractive crystals. A Bessel beam [26] with a concentric rings structure is a fundamental example of non-diffracting beams. The realization of the Bessel beam technique for holographic recording in the photorefractive crystals was reported in [27-31].

The cw single-mode laser at 532 nm and 140 mW maximum power was used in the experiments. For the formation of zero-order Bessel beams a single optical element, namely a conical lens-axicon, was used. The formation of Bessel-like photonic structures in doped LN crystals was performed by the scheme shown in Fig. 1. The crystal is positioned at the beam's overlapping zone half distance $Z_{max}/2$ which provides for the maximum number of Bessel rings. The beam size on the crystal input face equals ~ 3-4 mm. The inset in Fig. 1 shows the profile of the Bessel beam with a periodicity of 40 μm obtained by 178° axicon at the distance of $Z = 0.25$ m from the apex of the axicon.

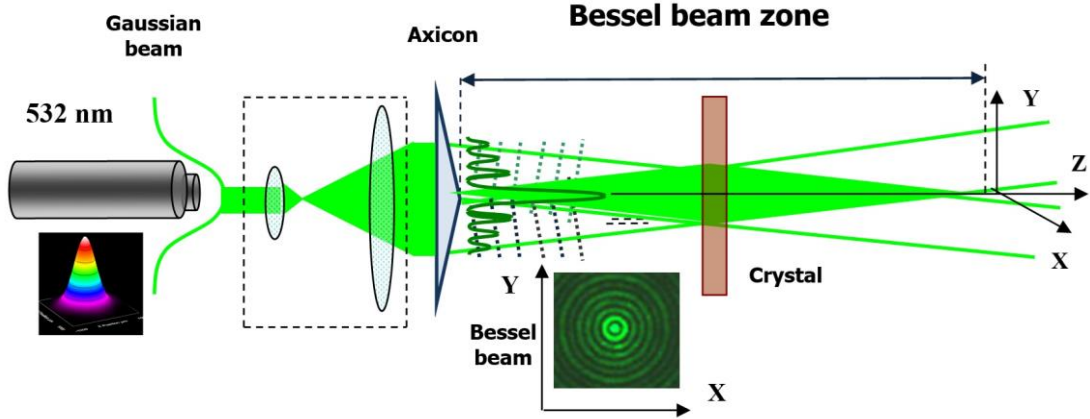

**Fig. 1.** Scheme of the experimental setup for the optical induction of a Bessel lattice in LN:Fe crystal. The insets show the Gaussian (left) and Bessel beam (right) intensity distributions.

(ii) *Lithium niobate crystal.* Among different photorefractive materials, the LN:Fe crystals are very promising for PV tweezers due to their excellent photorefractive properties. As a basic crystalline material a 0.05 wt % LN:Fe crystal with transverse (X×Y) size of 10÷20 mm and 1.5÷2 mm thickness along beam propagation Z is used. The used Y-cut crystals have an optical C-axis directed along the crystal surface. The value of the light-induced space-charged field $E \sim 10^7$ V/m is estimated in [11] for similar parameters of the experiment. The crystal allows also the total erasure of the recorded lattices by thermal heating to 150 °C providing a new recording. Thus, multiple uses of the same sample are possible making (DEP) forces geometry reconfigurable which is an important advantage.

(iii) *Sample preparation:* Ultra-pure double helix DNA from calf thymus (Sigma-Aldrich) is used in the experiment. The diameter of the DNA molecule is around a few nanometers. The length of the used DNA molecules is estimated at $L \sim 6.8$ μm (for ~20000 nitrogenous base pairs, with the distance between them ~0.34 nm). The long DNA in salt solution tends to roll into a random coil due to electrostatic repulsion in the phosphate groups.

The concentration of DNA 0.78 mg/ml base pairs (bp) Calf thymus DNA in $10^{-3}$ M NaCl solution is used. The NaCl solution is practically transparent in the range of spectrum 300–700 nm. DNA also is transparent in the visible range of the spectrum. The specific absorption peak of DNA is located at 260 nm wavelength. For fluorescent microscope measurements, the solution of 39.42 mg/ml ethidium bromide in $10^{-3}$ M NaCl was prepared. The measurements were carried out at the relative concentration of EtBr/DNA complexes $C_{EtBr}/C_{bp} = 0.5$.

(iv) *Detection.* The observation of refractive lattice recorded in LN:Fe crystal by 532 nm Bessel beam and DNA samples on the Bessel lattice refractive index pattern was performed by the Versamet Union interference contrast microscope and Nikon camera as well as Euromex iScope polarization microscope with VC.3042 camera and installed Image focus alfa software for images handling. The microscopes are operated in both reflection and transmission modes. The image mapping is performed for the operation of microscopes in transmission mode using objectives 5×, 10× and 20×.

The trapped DNA molecules are detected by the operation of the microscope in the transmission mode. A small drop (20-30 μl) of DNA in Na Cl buffer was placed on the LN:Fe crystal surface. A clean glass coverslip with a size matching the crystal size was gently placed over the drop. The prepared sample was observed by a phase microscope. The focal plane of the microscope objective was near the crystal surface. The depth of the focus of the microscope objectives, i.e. the distance over which the image plane can be displaced while the single object plane remains in a sharp focus, is a few micrometers. For this case, the trapped micro-object can be registered as refractive index non-uniformities on the Bessel lattice refractive index pattern ($n_e$ = 2.22 for LN:Fe) near the crystal surface in the same focal plane. For DNA, the refractive index has a value $n_{DNA} \approx 1.58$ in the visible range of the spectrum. Such observation allows simultaneous registration of both refractive lattice pattern and trapped micro-objects.

For observation by a fluorescent microscope, the DNA in salt solution (with the same 0.78 mg/ml concentration) stained by ethidium bromide was used. The drop of solution was placed on the glass substrate and the sapphire coverslip transparent for ultraviolet (UV) radiation was used. The sample was examined by fluorescence microscope (Labtron LFM B10) under UV-violet excitation. The digital registration was performed by a 16 MP Aptina CMOS sensor. For image processing the Allen Key software is used.

## 3. Experimental results

The imaging of DNA molecules in NaCl buffer was performed by PV trapping of the DNA molecules on the LN:Fe crystal surface and simultaneous observation of both the refractive lattice recorded in the crystal and the trapped molecules by phase-sensitive microscopy technique. This novel technique is compared with imaging of the DNA molecules by common fluorescence microscopy technique. In our experiments, the DNA molecules previously incubated with a selective DNA fluorescent label (ethidium bromide) are visualized by fluorescence under UV excitation. Experimental results are presented and compared to reveal the advantages and disadvantages of both methods.

### 3.1. Bessel-like photonic lattice in the Y-cut LN:Fe crystal

The recording of the Bessel lattice by a Bessel beam in the Y-cut 0.05 wt % LN:Fe crystal with optical C-axis directed along the crystal surface was performed for 30 min, which provided the saturation regime of recording. After switching off the laser beam the lattice recorded in the LN:Fe crystal was observed by an interference contrast phase microscope Versamet Union 2 operating in the transmission mode.

The fragment of the Bessel lattice with 40 μm periodicity is shown in Fig. 2(a). Fig. 2(b) shows the distribution of refractive index change $\delta n$ (in arbitrary units) of the Bessel lattice stripe along the diagonal radial direction $r$ marked by a rectangle. The sharp-peaked refractive index maxima are separated by wider refractive index minima with an average ratio of ~0.5 and a large depth of modulation.

While the Bessel rings have a nearly sinusoidal distribution except for a few central maxima, the space-charge field and photorefractive lattice do not replicate exactly the Bessel beam shape. It depends strongly on the modulation depth $m$. Large modulation depth of Bessel beam $m = 1$ leads to strong localization of the lattice fringes [12, 19, 21].

As the refractive index modulation is determined by space-charged PV field $E_{SC}$ ($\delta n_e$ ~0.5 $n_e^3$ $r_{33}$ $E_{SC}$, where $r_{33}$ is electro-optic coefficient [10, 11]), thus the $E_{SC}$-field should have a similar non-sinusoidal distribution. The lattice rings (Fig. 2(a)) appear more uniform compared with the actual Bessel beam profile (Fig. 1) as a result of saturation of space charge-field during 30 min recording.

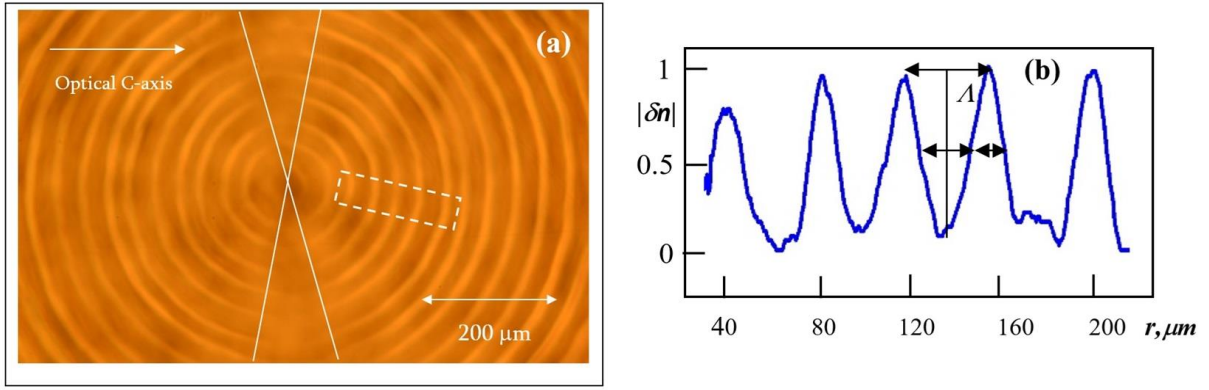

**Fig. 2.** (a) Phase microscope image of the Bessel lattice with a periodicity of 40 μm recorded by 532 nm and 40 mW power Bessel beam during 30 min in LN:Fe crystal. The crossed lines mark the areas where the lattice is recorded with high contrast (left and right sectors) and are not recorded at all (upper and lower sectors) due to the annular symmetry of the Bessel beam crystal and the anisotropy of the crystal. The horizontal arrow shows the direction of the C-axis of the crystal. (b) Distribution of refractive index change $\delta n$ (in arbitrary units) of the Bessel lattice stripe along the diagonal radial direction $r$. The arrows show the period of the lattice $\Lambda = 40$ μm.

We also checked that after four months of storage of the LN:Fe crystal with recorded Bessel lattice in the dark, the lattice is still observable by phase microscope with practically unchanged contrast and thus retains the ability to trap the micro- and nanoparticles.

### 3.2. PV trapping of DNA molecules in NaCl solution on the LN:Fe crystal surface and their imaging

The following technique was used for the observation of DNA in NaCl buffer by phase microscope.

A small drop of a few μL was placed on the clean LN:Fe crystal in the area of the recorded hologram. The drop was gently pressed down by a clean glass plate with a thickness of 400 μm and transverse size equal to the crystal size of 10.1×10.2 mm$^2$. The NaCl film with dispersed DNA in the prepared sandwich had around 10 μm thickness and its volume is estimated at 10$^{-6}$ dm$^3$ = 1 μL. The concentration of DNA in NaCl buffer was chosen at 0.78 mg/ml, taking into account the previous finding [21] that for the concentration of 0.58 mg/ml, numerous DNA molecules are detected on the crystal surface in the area of the recorded hologram.

The observation of the Bessel refractive lattice and trapped molecules was performed by a phase microscope operating in the transmission mode. Such observation provided the simultaneous observation of both DNA molecule and Bessel refractive lattice. The DNA molecules with $n_{DNA} \approx 1.58$ in the visible range of the spectrum are registered as refractive index nonuniformities on the Bessel lattice refractive index pattern in LN:Fe crystal ($n_e \approx 2.22$).

Numerous measurements are made to obtain the statistics of the disposition of the DNA molecules relative to the lattice fringes.

The microscope image of the distribution of DNA molecules in NaCl solution on the crystal surface in the area where the Bessel lattice is recorded, for a few minutes delay after sandwich preparation, is shown in Fig. 3. The microscope image demonstrates an irregular disposition of DNA molecules relative to the refractive grating fringes. Some of the particles disposed on the lattice maxima are depicted by solid circles, while the dashed circles show the particles located in the minima of the lattice. Some of the particles depicted by ovals are out of the focal plane of the lattice fringes.

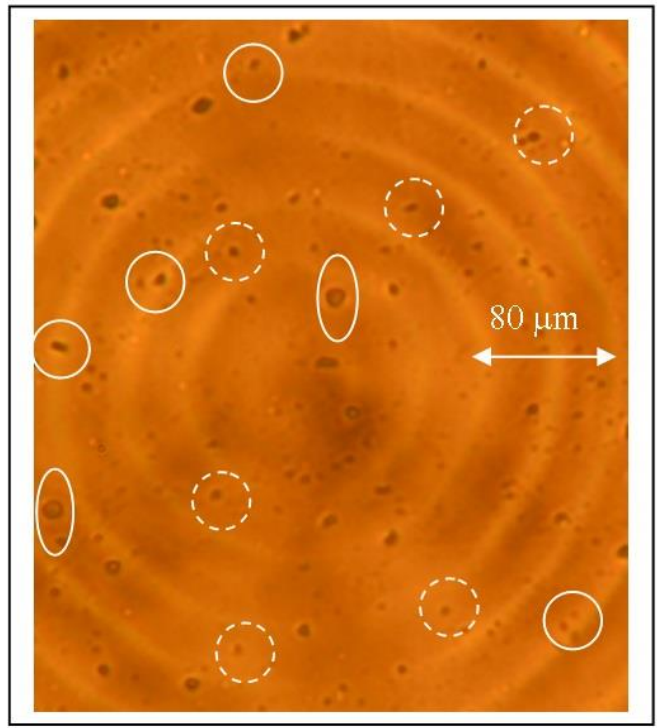

**Fig. 3.** The distribution of DNA molecules in NaCl solution with 0.78 mg/ml concentration on the surface of LN:Fe crystal in the area with the recorded Bessel lattice, observed by phase microscope just after preparation of thin film of NaCl with dispersed DNA. The microscope image shows the irregular disposition of the DNA molecules relative to the Bessel lattice fringes. Some of the particles disposed on the lattice maxima are depicted by solid circles. The dashed circles show the particles located in the minima of the lattice. Some of the particles depicted by ovals are out of the focal plane of the lattice fringes.

Observations performed after 30 min delay show that the DNA molecules are immobilized and resided on the crystal surface. They appear as sharp black spots on the Bessel lattice pattern. Results of phase microscope observations of DNA trapping are shown in Figs. 4(a) – 4(j) and demonstrate certain regularity. DNA molecules appear as black spots at the borderlines of the Bessel lattice rings. Some particles are depicted by circles. It is worth noting that particles are located at both the inner and outer sides of the lattice rings, which are shown by ovals in Fig. 4(a) and enlarged Figs. 4(b), 4(c) and 4(g). Among the 40 particles shown in Fig. 4(a), only four exceptional cases of DNA localizations in the minima of lattice fringes are observed. Two of them are marked in Figs. 4(a) and 4(e) by a rectangle. In general, such deviation can be connected with local defects in the crystal and irregularity of the PV field.

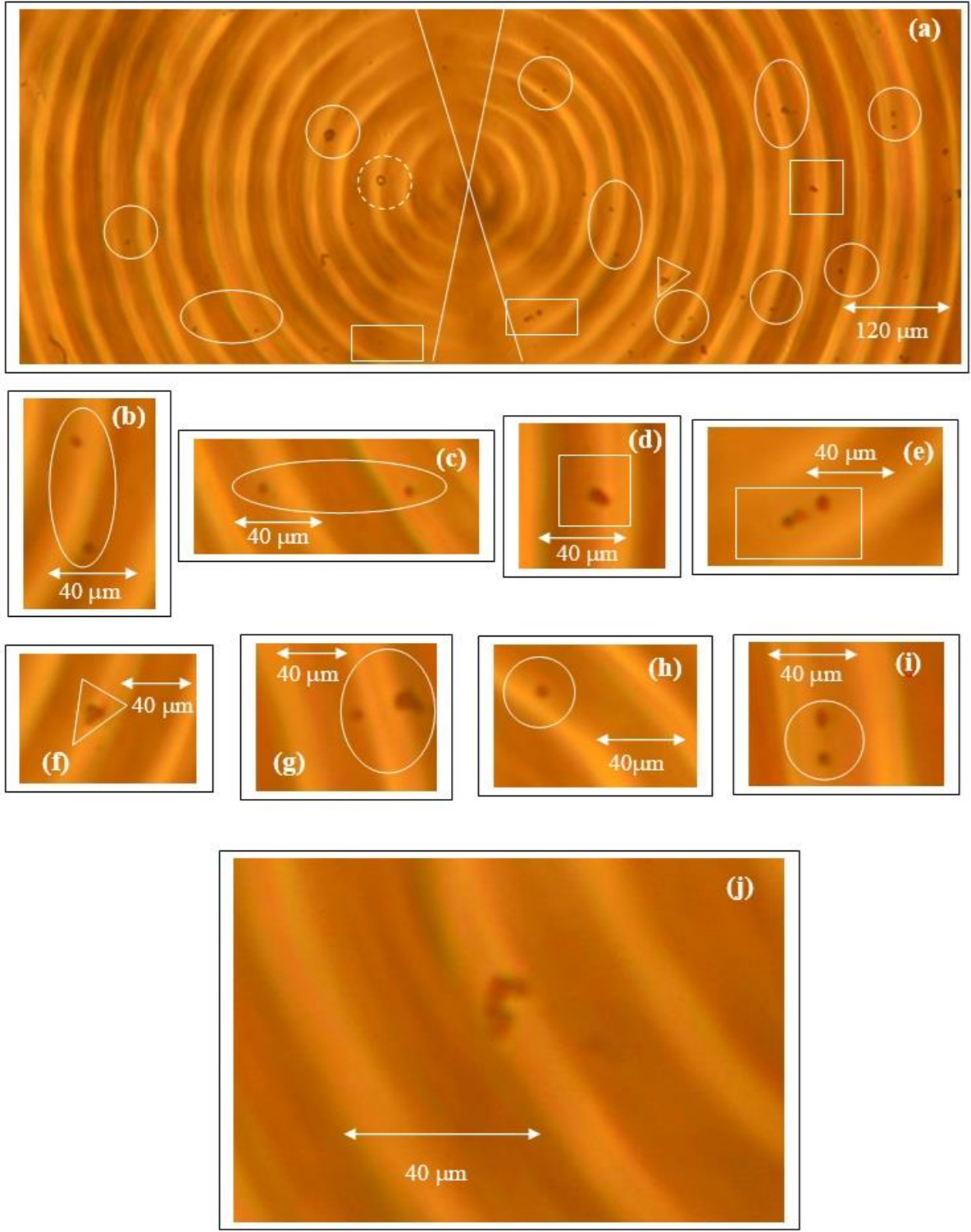

**Fig. 4.** (a) Phase microscope image of the trapped DNA molecules in NaCl solution and their disposition relative to the Bessel lattice rings for 30 min delay. (b) - (i) Enlarged patterns of some sections selected in (b) by solid circles, ovals, a square, rectangles, and a triangle. (j) The arc-shaped elongated DNA with a length of 14 μm. Explanations are given in the text. The dotted white circle in (a) depicts the folded DNA molecule which is out of the focal plane of the objective and appears as an open circle.

DNA molecules have mainly globular shape with an average size of 4 μm (Fig. 4(h)). However, the larger size elongated (Figs. 4(d) and 4(i)) or triangle shape (Figs. 4(f) and 4(g)) folded molecules also are observed. An interesting observation is the formation of elongated arc-shaped DNA with a thickness of 4 μm and a total length of 14 μm (Fig. 4(j)). The histogram of the number of nearly globular particles depending on their sizes in the Bessel lattice area is shown in Fig. 5. The most probable size (MPS) of the folded DNA molecules is measured at 4 μm. The DNA observations also were performed by polarization microscope Euromex iScope and similar results were obtained.

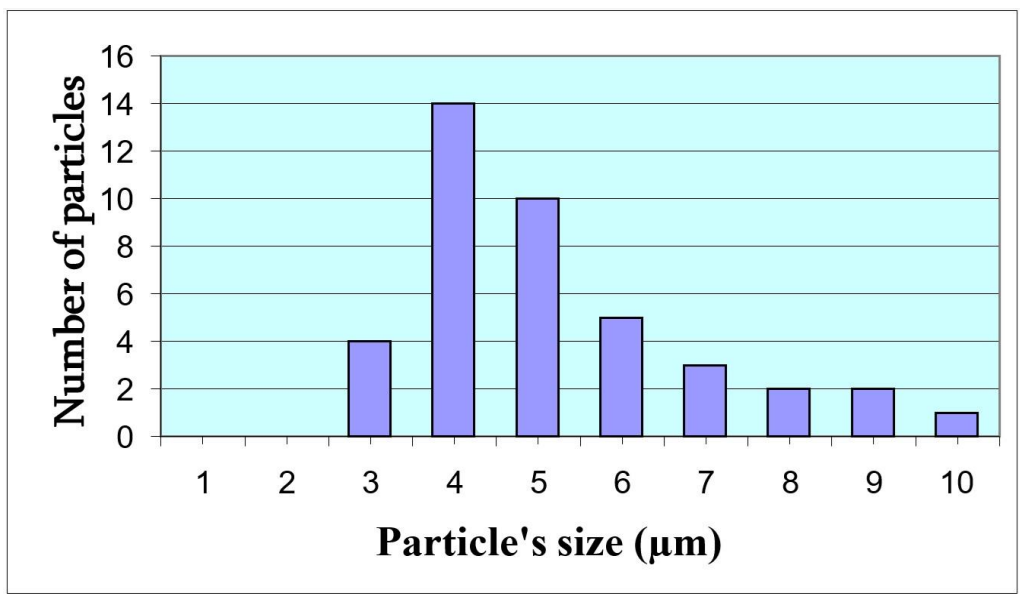

**Fig. 5.** The distribution of the number of trapped particles depending on their size in the area of the Bessel lattice on the crystal surface.

The simultaneous observation of the refractive lattice rings and trapped objects is a great advantage of the suggested microscopy method based on PV tweezers. In addition to accurate measurement of the size and shape of the trapped particles, the method also allows the determination of the charged or neutral nature of the trapped objects by their disposition relative to the lattice fringes. This possibility will be discussed in detail in the Section 5, Discussion.

*3.3 Imaging of DNA molecules by fluorescence microscope*

Commonly the "folded" DNA molecules are studied by fluorescence microscopy technique [22]. The DNA labeled by dye molecules is visualized under UV excitation.

The ethidium bromide (EtBr) is widely used for labeling DNA molecules. EtBr when bound to double-strand DNA between base pairs exhibits the increased fluorescence. EtBr has an excitation peak at 300 nm and fluoresces red-orange color with a peak at 603 nm wavelength (see [32] and Refs. therein).

We performed comparative studies of the DNA in the NaCl salt solution by standard fluorescence microscopy technique. In this experiment, the ethidium bromide with a concentration of 39.42 mg/ml in $10^{-3}$ M NaCl was added in the small portion of the stock salt solution of 0.78 mg/ml base pairs calf thymus DNA in $10^{-3}$ M NaCl, which was previously used in the series of measurements with trapping and phase-sensitive imaging of DNA. The relative concentration of EtBr/DNA complexes is CEtBr/Cbp = 0.5. DNA imaging by fluorescence microscope was performed without any trapping, as it is commonly used in such experiments.

A small drop (5-10 μl) of the solution was placed on the glass substrate. The sapphire coverslip with 400 μm thickness, transparent for ultraviolet (UV) radiation was gently placed over the drop. Observations were performed for a 10-minute delay after switching on the UV excitation to have stable illumination. Figs. 6(a) - 6(e) show some examples of the fluorescence microscope images at different areas of DNA solution spread over the glass substrate. The detected DNA molecules have different shapes, namely nearly globular (depicted by solid circles in Figs. 6(a) and 6(b)), elongated (shown by ovals in Fig. 6(a)), ark-shape (Fig. 6(c), triangle-shape (Fig. 6(d)), and nearly open circle shape (marked by a dashed circle in Figs. 6(a) and 6(e)).

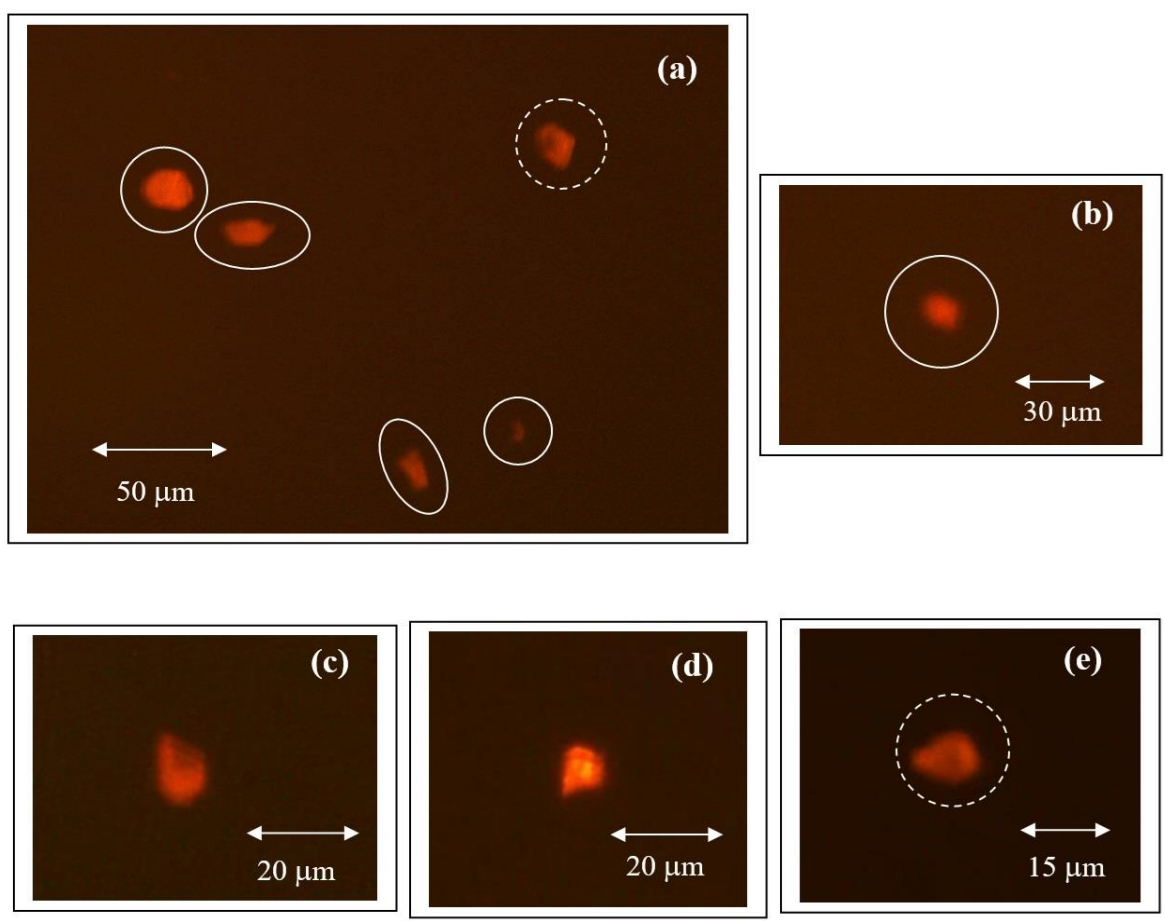

**Fig. 6.** Examples of the fluorescence microscope images of DNA molecules stained by ethidium bromide at different areas of DNA solution spread over the glass substrate. The detected DNA molecules have different shapes, namely nearly globular depicted by solid circles in (a) and (b), elongated shown by ovals in (a), ark-shape (c), triangle-shape (d)), and with the tendency to form an open circles depicted by dotted white circles in (a) and (e). The patterns are obtained with 10× magnification of the objective. Explanations are given in the text.

The histogram of the number of fluorescent DNA molecules depending on their sizes in the area of observation is shown in Fig. 7. Total number of processed nearly globular DNA molecules is 44. The lowest size of the folded DNA with globular shape detected by microscope for the prepared DNA sample equals 4 μm. The MPS of the folded DNA molecules is measured at ~10 μm.

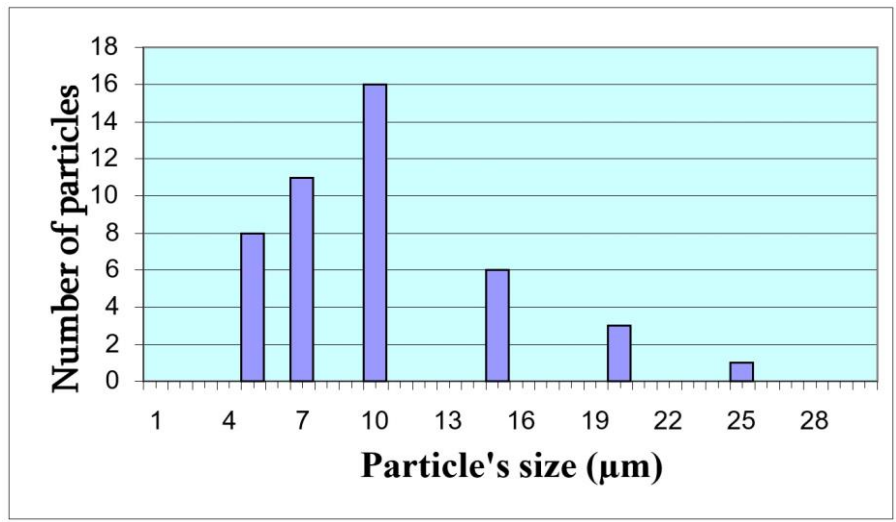

**Fig. 7.** The histogram of the number of fluorescent DNA molecules depending on their sizes in the area of observation.

Similar DNA shapes and sizes were observed by fluorescence microscopy, for example, in [22].

## 4. Discussion
### *4.1. Optical generation of PV fields inside the crystal and near its surface and force map*

Space-charge PV fields induced in the photorefractive crystals by inhomogeneous illumination modulate the refractive index via the electro-optic effect thus creating the refractive structures inside the crystal. The contribution from the isotropic diffusion effect is one order of magnitude smaller. The light-induced refractive structure in the crystal nearly replicates the illumination pattern. The recording of the refractive lattice by the Bessel beam is described in detail in [19].

The light-induced PV field not only affects the refractive index inside the crystal but also serves as a source of periodically distributed electric fields near the crystal surface. Since for Y-cut LN:Fe crystal the optical C-axis is directed along the crystal surface, the alternating PV field also is periodically distributed along the crystal surface as it is shown in Fig. 8(a). The PV field near the surface of the crystal is evanescent and falls rapidly with a distance z from the crystal surface as $\exp(-3z/\Lambda)$, where $\Lambda$ is the period of the lattice [12, 13]. Thus, PV field can act on micro- or nano-objects when they are close enough to the crystal surface. The PV fields $E$ induce electrical forces on either the particles that have electrical charge $q$ (electrophoretic forces $F_{EP} = qE$) or the neutral particles that become polarized by an electric field's DEP forces.

Thus, the PV tweezers can work as a charge sensor. For example, for the negatively charged particles, the electrophoretic forces will move particles to the area of the light-induced alternating PV field which has a positive sign, and vice versa. The neutral particles became polarized by a nonuniform electric field and experience DEP forces that depend on the particle volume and the gradient of *E*-field intensity (see formula (1)). The force sign, indicating its repulsive or attractive nature, depends on the Clausius - Mossotti factor *K*.

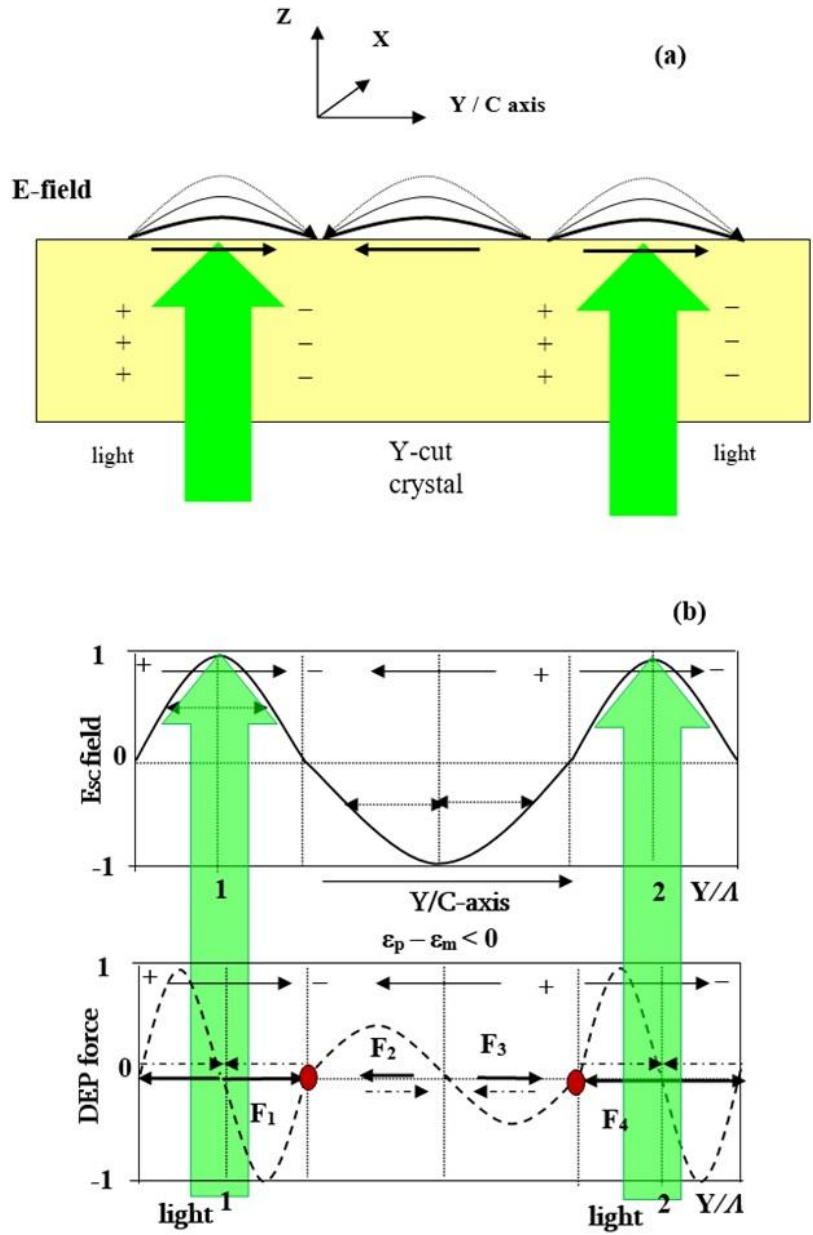

**Fig. 8.** (a) The intensity-modulated light induces an alternating photovoltaic field that is directed along the C-axis of the crystal parallel to the crystal surface for Y-cut LN:Fe. The photovoltaic field out of the crystal near its surface is evanescent and falls rapidly with a distance z from the crystal surface as exp ( −3z/Λ), where Λ is the period of the lattice [12,13]. (b) Schematic of Bessel beam-induced non-sinusoidal space charge field $E_{SC}$ pattern (solid curve) and DEP force $F_{DEP}$ (dotted curve) distribution on the Y-cut crystal surface versus spatial coordinate Y in the units Y/*Λ*, where *Λ* is period of the lattice. The thin arrows show the directions of the $E_{SC}$ field (positive and negative) along the crystal surface. The solid arrows show the directions of the repulsive DEP forces for $\varepsilon_p - \varepsilon_m < 0$. DEP forces move the micro-objects shown by red spots toward zeros of the space-charge field thus trapping them near the borderlines of lattice rings under the combined action of the forces F1- F2 and F3 - F4. For comparison, the dash-dot arrows show the directions of attractive forces for $\varepsilon_p - \varepsilon_m > 0$ Areas of the crystal illuminated by green light are shown by vertical arrows.

For sinusoidal grating and attractive forces ($\varepsilon_p > \varepsilon_m$) the dielectric particles will be trapped in the areas of PV field maxima and minima, i.e. in the areas of refractive grating maxima and minima, where electric field gradient and DEP force are zero. For non-sinusoidal grating recorded by a Bessel beam, the DEP force dominates in the first positive half period of the sharp E-field due to a large E-field gradient. Thus, for relatively massive particles, they will be trapped in the sharp-peaked maxima of refractive grating. Such behavior was observed for the trapping of $CaCO_3$ microparticles in the maxima of the Bessel lattice recorded in LN:Fe crystal [19].

For metal particles $\varepsilon_p$ is negative thus ($\varepsilon_p - \varepsilon_m$) < 0 and if ($\varepsilon_p + 2\varepsilon_m$) > 0 the force is repealing. For this case, the E-field and DEP force distribution provides the localization of micro/nanoparticles at the borderlines of the lattice rings. Such an experiment was performed in [19] for Ag particles in glycerin suspension. The graphical analysis of repulsive forces for trapping of Ag particles is given in [19].

The PV tweezers approach was successfully extended to the study of DNA in NaCl solution [21]. While the individual DNA molecules have a negative charge, the PV trapping approach allowed the revealing of the formation of the neutral DNA microparticles. This fact was detected by the disposition of DNA molecules at both extremes of the lattice rings where the PV field has positive and negative signs, respectively. The physical mechanism is the screening by $Na^+$ counterions available due to the dissociation of NaCl in the water.

However, the comparative studies of the imaging of the folded DNA molecules by the elaborated PV trapping technique and the state-of-the-art method by fluorescence microscopy technique [22] remained open. These actual studies were performed in this work. In the next section, the advantages and disadvantages of DNA imaging by phase-sensitive and fluorescence techniques are discussed.

*4.2. Comparative studies of DNA imaging by phase-sensitive and fluorescence microscopy*

The PV tweezers are exploited for the imaging of the DNA molecules trapped on the crystal surface and their sizing with the use of phase microscopy. The advantage of the novel method is the non-invasive phase-sensitive imaging of DNA molecules and accurate measurement of both the size and shape of DNA molecules. This method also provides simultaneous observation of the recorded inside the crystal refractive lattice (PV field configuration) and the trapped micro-objects on the crystal surface by the phase microscope. This, in turn, allows the determination of the charged or neutral nature of the trapped bio-objects by their disposition relative to the lattice fringes.

The results of imaging of DNA molecules, shown in Fig. 4, demonstrate the preferential formation of globular-shaped DNA molecules with the most probable size of 4 µm. Fig. 5 shows the results of statistical analysis of the number of particles depending on their sizes in the Bessel lattice area. Among the 40 DNA molecules shown in Fig. 4(a), most of them have globular shape. This result also is confirmed by many other series of measurements.

The observation of DNA stained with EtBr molecules by fluorescent microscope showed various tertiary configurations adopted by the folded DNA molecules. Moreover, while the sizes of "pure" DNA molecules vary in the range of 3-10 µm (Fig. 5) with MPS = 4 µm, the dye-labeled DNA molecules have sizes in the range of 4 - 25 µm with MPS = 10 µm (Fig. 7). This is the result of the alteration of the structure, shape, and size of the labeled

DNA by dye molecules. We can conclude that the size of folded DNA molecules stained with dye molecules is 2.5 times larger compared with the "pure" DNA.

While the individual DNA molecules have a negative charge [33], the aggregation [34] and/or condensation (see [35] and Refs. therein) of individual DNA molecules to the larger-size neutral clusters via the positive metal ions available in a salt solution ($Na^+$ in NaCl solution) are possible. The concentration of 0.78 mg/ml of DNA molecules in $10^{-3}$ M NaCl solution provides nearly equal concentrations of DNA nucleotides and $Na^+$ counterions with the order of magnitude of $10^{-3}$ M.

The analysis of the disposition of the folded DNA molecules relative to the lattice fringes shows that the trapped DNA clusters are mainly localized at the borderlines of the lattice rings, i.e. on both positively and negatively charged extremes of the lattice rings. It means that for the condition of our experiment, the DEP forces influence the neutral objects. Fig. 8(b) shows the alternating space-charge fields on the LN:Fe crystal surface and the schematic pattern for the distribution of the PV field (upper figure) and DEP forces (lower figure), respectively. The DNA molecules ($\varepsilon_p = 8.3$ at 532 nm [36]) dispersed in $10^{-3}$ M NaCl solution ($\varepsilon_m \sim 80$) experience the DEP forces which, according to the formulae (1), are repulsive. The repulsive forces $F_1 - F_2$ and $F_3 - F_4$ lead to the trapping of DNA molecules at the borderlines of lattice rings according to our experimental results (Figs. 4(b) – 4(i)).

The length of the used DNA molecules is estimated at $L \sim 6.8$ μm (~20000 nitrogenous base pairs, with the distance between them ~0.34 nm). The long DNA in salt solution tends to roll into a random coil due to electrostatic repulsion in the phosphate groups. The "folded" DNA has the characteristic persistent segments of the length $l \sim 100$ nm [37]. The mean diameter $D$ of the DNA coil is estimated at ~0.8 μm with the use of a simple formula $<D^2> = Ll$ [38].

However, the imaging of the trapped DNA molecules by phase-sensitive microscopy showed that the sizes of the folded DNA molecules vary in the wide range from 3 μm to 10 μm with the most probable size of 4 μm (Fig. 5). The detection of larger-size folded DNA confirms the concept of aggregation of individual DNA molecules to neutral clusters. The DNA clusters are nearly globular, elongated, triangle-shaped, and even extra-long 14 μm arc-shaped (Fig. 4). The specific elongated arc-shaped DNA is settled on the maximum of lattice fringe, while the edges of the arc are localized at the borderlines of the lattice ring (Fig. 4(j)). Such disposition is not in contradiction with the neutral nature of the DNA cluster. Similar arc-shaped-folded DNA also was observed in [22] as well as in our fluorescence microscope measurements (Fig.6(c)).

The trapping of most of the DNA molecules on both extremes of lattice rings is an indication that we deal with neutral clusters that experience PV-induced DEP forces.

The comparison of both methods, namely phase-sensitive and fluorescence, shows the advantage of the PV-tweezers-based phase-sensitive technique. The method is non-invasive, as it does not require the labeling of DNA molecules and provides additional information on the charged or neutral nature of the trapped bio-objects by their disposition relative to the lattice fringes.

## 5. Conclusion

A novel *non-invasive* microscopy technique for imaging and sizing of folded DNA molecules by PV tweezers and phase-sensitive detection is elaborated and realized. This leads to the novel optical microscopy technique with inserted PV tweezers.

The novel microscopy concept is successfully applied for the manipulation, trapping, imaging, and sizing of DNA molecules in NaCl solution.

The PV- tweezers, which are the basis of the suggested microscopy technique, have the following important parameters; they are the chip-scale, 2D, high efficiency, with operation in an autonomous regime (lab-on-a-chip). The suggested novel method provides the imaging and accurate measurement of both the size and shape of various bio-objects, as well as the determination of the charged or neutral nature of the trapped bio-objects by their disposition relative to the lattice fringes.

The comparative study of the imaging of the folded DNA molecules by the elaborated non-invasive phase sensitive PV trapping technique and by commonly used fluorescence microscopy technique is performed, which demonstrates that labeling of DNA molecules results in overestimation of the molecular size.

The application of the elaborated novel non-invasive microscopy technique to the other bio-objects is foreseen.

**Acknowledgements**

The authors are grateful to Dr. Edvard Kokanyan for providing the lithium niobate crystals in the framework of the International Science and Technology Center Grant, Project A-2130. The authors express sincere gratitude to Dr. Lusine Aloyan for the bio-samples preparation and discussions.

**Funding**

The presented study is performed with the financial support of the Higher Education and Science Committee of the Ministry of Education, Science, Culture and Sports of RA, in the frames of project N1-6/24-I/IPR.

**CRediT authorship contribution statement**

Lusine Tsarukyan: Investigation, Analysis. Anahit Badalyan: Investigation, Analysis. Rafael Drampyan: Supervision, Conceptualization, Investigation.

**Declaration of Competing Interest**

The authors declare that they have no known competing financial interests or personal relationships that could have appeared to influence the work reported in this paper. No potential conflict of interest was reported by the authors.

**Data availability**

Data are available upon reasonable request to the corresponding author.